\documentstyle[12pt]{article}
\newcommand{\p}{$\overline{p}$}

\def\p{$\pi$ }
\def\r{$\rho$ }
\def\o{$\omega$ }
\def\s{$\sigma$ }
\def\e{$\eta$ }

\def\ppp{$pp \rightarrow pp \pi^0$ }

\def\fm3{fm$^3$}
\def\fmm3{fm$^{-3}$}
\def\fsi{FSI\ }

\def\s11{S11\ }

\begin{document}
\title{\hspace{3.0in}{\small BGU PH-96/06}\\
\vspace{0.8cm}
S-wave $\pi^0$ production In pp Collision In A Covariant
OBE Model}
\author{ E. Gedalin, A. Moalem and L. Razdolskaja\\
        {\small  Department of Physics, Ben Gurion University, 84105
Beer Sheva, Israel}}
\maketitle
\begin{abstract}
The total cross section for the $pp \to pp \pi^0$ reaction at energies 
close to threshold is calculated in a covariant one-boson-exchange 
model. The amplitudes for the elementary $B N \to N \pi^0$ processes
are taken to be the sum of  s, u and t pole terms. 
The main contributions to the primary production
amplitude is due to a $\sigma$ meson exchange, which is strongly enhanced 
due to the t pole term. 
The NN and $\pi$N final state interactions are included coherently.
The effects due to the $\pi$p interactions in pure isospin $\frac {1}{2}$
and $\frac{3}{2}$ channels are sizable but when taken in the appropriate
isospin combination almost cancel out.
Both the scale and
energy dependence of the cross section are perfectly reproduced.
\end{abstract}
\vspace{2.0 cm}
\begin{center}
(To be submitted for publication in Phys. Lett. {\bf B})\\
\end{center}
\vspace{2.0 cm}

PACS number(s) : 13.75.Cs, 14.40.Aq, 25.40.Ep
\newpage

In a recent study of the \ppp reaction at energies close to
threshold, it is found that the angular distributions of the outgoing
particles are isotropic, in agreement with the assumption that the
reaction proceeds as a $^{33}P_0 \to {}^{31}S_0 s_0$ 
transition\cite{bondar,meyer}.
Further more, the energy dependence of the total cross section, particle 
spectra and angular distributions are well reproduced by taking into
account the pp final state interactions (FSI) only. The $\pi$N interactions
seem to play a minor role, in marked difference with the similar 
$pp \to pp \eta$ reaction, where the $\eta$N force is found to be essential
for explaining the energy dependence of the cross section
\cite{moalem1,moalem2}.
Several model calculations\cite{koltun,miler,niskanen} 
of S-wave pion production, which are based on
a single nucleon and a pion rescattering mechanism, 
under estimate the cross section by a factor of 3-5. 
Inspired by a study of $\beta$-decay in nuclei, which indicates that
the axial charge of the nuclear system is enhanced by heavy meson
exchanges and, by using a simple operator form of the NN potential,
Lee and Riska\cite{lee} have shown that meson exchange currents could 
explain the scale of the cross section.
Horowitz et al.\cite{horowitz} performed similar calculations
based on explicit one-boson-exchange (OBE) model 
for the NN interaction as well
as for the evaluation of meson exchange contributions.
The evaluation of these contributions depends on the virtuality of the 
meson exchanged and on how the on mass-shell amplitudes are extended into
the off mass-shell region. There are several approach based on field 
theoretical models which allow such extension to be made but the results 
are model dependent. In the traditional phenomenological 
treatment\cite{koltun,laget,miler,niskanen}, the
off mass-shell and on mass-shell amplitudes are of the same order of
magnitude. Recently, several groups\cite{hernandez,hanhart,park,miller}
have concluded that the off mass-shell $\pi$N rescattering amplitude is
enhanced with respect to the on mass-shell amplitude. 
Hernandez and Oset\cite{hernandez} 
by applying current algebra and PCAC constraints argue that this enhancement
may bring the calculated cross section for the \ppp reaction 
into agreement with experiment.
More detailed momentum-space calculations by Hanhart et al.\cite{hanhart}
confirm this enhancement of the off mass-shell $\pi$N 
scattering amplitude but conclude that
it is still too short to explain the scale of the cross section. 
Park et al.\cite{park} and Cohen et al.\cite{miller}
have applied a chiral perturbation theory, including
chiral order 0 and 1 Lagrangian terms. They have shown that the off mass-shell
$\pi$N scattering amplitude is enhanced considerably but has an opposite
sign with respect to the on mass-shell amplitude. Because of this 
difference in sign, the rescattering term and the Born term contributing
to the \ppp reaction interfere destructively, making the theoretical
cross sections much smaller than experimental values. 
This enforces  the importance 
of the heavy meson exchange contributions. Part of these contributions 
was included in Ref.\cite{miller}.

In the present work we propose a covariant OBE model based on a two-step
mechanism, where a virtual boson 
B (B=\p , $\sigma$ , \e , \r , \o ...) created on
one of the incoming nucleons, is converted into a $\pi^0$ 
meson on a the second
via a $BN \to \pi^0 N$ process (see diagram of Fig. 1). 
Similar to $\pi^0$ electroproduction
amplitudes, 
the amplitudes for the elementary 
$B N \to N \pi^0$ processes are taken to be the sum of s, u and t
pole terms. In fact only for the "effective" $\sigma$ meson, a t pole 
term does contribute, 
giving rise to a strong enhancement of the cross section. 
With this model we obtain perfect agreement
with the empirical cross-section data. 
Though differing in many details, the present work agrees
with Horowitz et al.\cite{horowitz},
that the $\sigma$ meson exchange plays an important role in  pion
production near threshold. Our model however is a natural extension of the
relativistic OBE model used to analyze NN scattering data\cite{machleit},
where a single nucleon mechanism is not possible and all contributions 
correspond to completely connected diagrams. In the present formulation
of the model, we may use any form of the $BN \to \pi^0 N$ amplitude 
so that comparison with other model calculations can be performed 
directly.

We use the following Lagrangian interaction: 
\begin{eqnarray}
L & = & i g_{\pi NN} \bar{N} \gamma^5 {\bf \tau} N {\bf \pi} +
i g_{\eta NN} \bar{N} \gamma^5  N \eta + 
g_{\sigma NN} \bar{N}  N {\bf \sigma} +  \nonumber \\
  &   & \bar{N}\left( g_{\rho NN} \gamma^{\mu} + i \frac {f_{\rho NN}}{2M}
\sigma^{\mu \nu} q_{\nu}\right)
{\bf \tau} N {\bf \rho}_{\mu} +
g_{\omega NN} \bar{N} \gamma^{\mu}  N {\bf \omega}_{\mu} +
g_{\delta NN} \bar{N}{\bf \tau}  N {\bf \delta} \ \ ,
\end{eqnarray}
with obvious notation. The coupling constants and the meson masses are
taken from a fit to NN scattering data and are listed in table 5 of 
Ref.\cite{machleit}. Using pseudovector couplings for the pseudoscalar
mesons in the expression above would not alter the principal conclusions 
from the present work.

Following  Refs.\cite{moalem1,moalem2} 
we write the transition amplitude in the form,
\begin{equation}
T_{23} \approx M^{(in)}_{23} \ T^{(el)}_{33}\ \ ,
\label{eq:2}
\end{equation}
where $M^{(in)}_{23}$ is the primary production amplitude which describes
the transition from a two-nucleon state  to a three-body state of two nucleon 
and a neutral pion, and $T^{(el)}_{33}$
is a \fsi correction factor taken to be the on mass-shell elastic
scattering amplitude of the $pp \pi^0 \to pp \pi^0$ transition.
The validity of this approximation for \p and \e meson production in NN
collisions is discussed in length in Refs.\cite{moalem1,moalem2}.
Here we note that :
(1) all inelastic interactions ( and hence the coupling to other channels )
are included into $M^{(in)}$ while $T^{(el)}_{33}$ depends on elastic
interactions among the reaction products only, 
(2) as in the coherent formalism for three-body processes\cite{amado,cahill}, 
the different two-body interactions in the final state
contribute coherently and (3) the transition amplitude has an overall
phase factor identical to that of $T^{(el)}_{33}$ as required by general
quantum mechanics constraints\cite{amado}.

To calculate the transition amplitude we apply covariant 
perturbation theory techniques. 
We assume that the reaction is dominated by the mechanism depicted in Fig. 1
and write $M^{(in)}$ in the form, 
\begin{equation}
M^{(in)} = \sum_{B}^{} \left[ T_{BN \to \pi^0 N} (p_4,k;p_2,q) G_B (q_{})
S_{BNN} (p_3,p_1)\right] + [1 \leftrightarrow 2 ; 3 \leftrightarrow 4]\ \ ,
\label{eq:3}
\end{equation} 
where $p_i$, $q_{}$ and $k$ are 4-momenta of the ith nucleon, 
the exchanged boson 
and the outgoing pion. The sum runs over all possible B boson exchanges
that may contribute to the process. The
bracket $[1 \leftrightarrow 2 ; 3 \leftrightarrow 4]$ stands for a similar
sum with the $p_1$, $p_3$ and $p_2$, $p_4$  momenta interchanged. 
In Eqn. 3, $T_{BN \to \pi^0 N}$ is the  amplitude for the 
$BN \to \pi^0 N$ transition, $G_B (q_{})$ and $S_{BNN} (p_3,p_1)$ are
the propagator and source function of the meson exchanged, respectively.
By using covariant meson propagators and 
form factors and covariant parametrizations for $S_{BNN}$ and 
$T_{BN \to \pi^0 N}$, the transition amplitude is 
expressed in terms of invariant functions.
For a scalar and pseudoscalar meson exchanges,
\begin{eqnarray}
S_{SNN} (p_1,p_3) = \bar {u} (p_3)\  I\  u (p_1)\  F_S (q_{})\ \ ;
S_{PNN} (p_1,p_3) = \bar {u} (p_3)\  \gamma^5\  I\  u (p_1)\  F_P (q_{})\ \ ,
\label{eq:4}
\end{eqnarray} 
where $F_B(q_{})$ is a source form factor and
$I$ the appropriate isospin operator. Here $u$ is a nucleon 
Dirac spinor and $p_3 = p_1 - q_{}$ is the final nucleon momentum. A
most general vector source contains vector and  tensor current terms,
\begin{equation}
S^{\mu}_{VNN} (p_1,p_3) = \bar {u} (p_3)\left[ \gamma^{\mu} F^{(1)}_V (q^2_{13})
   +  i \sigma^{\mu \nu} q_{\nu} F^{(2)}_V (q^2_{13}) \right] I u (p_1) \ \ ,
\label{eq:5}
\end{equation} 
with the vector source form factors, $F^{(i)}_V$, being analogous to the  
electromagnetic form factors of the nucleon. In the calculations 
to be presented
below, all source form factors are taken in their lowest order perturbative
approximation.
The expression, Eqn. 5, contains conserved currents only
and therefore satisfies current conservation.

The amplitude for a scalar meson-nucleon scattering, 
$S N \to P N$, is written in 
the usual form\cite{gaziorovitch},
\begin{equation}
T_{S N \to P_2 N} (p_4, k; p_2, q) = \bar {u} (p_4) \gamma^5\ \left[ A + 
\frac {1}{2} ( k\! \! \! /  +  q\! \! \! /  ) B\right] u (p_2) \ \ ,
\label{eq:7}
\end{equation} 
where $A$ and $B$ are functions of the Mandelstam variables, having the
same isospin structure. To obtain the amplitude corresponding to
a to a pseudoscalar meson scattering, $P_1 N \to P_2 N$, one has to remove 
the $\gamma^5$ in the expression above.

Similarly, for a vector meson 
exchange , the amplitude for the $ V N \to \pi^0 N $ transition is, 
\begin{equation}
T^{\mu}_{VN \to \pi^0 N} (p_4, k; p_2, q)  =  \bar{u} (p_4)\gamma^{5} \left[
\gamma^{\mu}  {\cal A}_{12}
+ p^{\mu}_{4}   {\cal A}_{34} + k^{\mu}  {\cal A}_{56}
 + q^{\mu} {\cal A}_{78}\right] u(p_2) \ \ ,
\label{eq:8}
\end{equation} 
where ${\cal A}_{ij}$ stands for the combination 
$(A_i\  +\  k\! \! \! /  A_j)$.

In the analysis to be presented below, we include \p , $\sigma$ , \e , \r and \o
meson exchanges. In order to calculate these contributions to the 
transition amplitude for the \ppp reaction we call attention to certain
kinematical features of this process.

One important aspect of the mechanism of Fig. 1 is that at the $\pi^0$
production threshold, the transferred 
4-momentum is space-like,  $q^2  = -3.3 fm^{-2}$. 
This is very much the same kinematics that occurs in $\pi^0$ electroproduction
through vector meson exchanges, suggesting that the half off mass-shell 
amplitudes $T_{BN \to \pi^0 N}$ which appear in Eqn. 3, can be calculated
using the usual calculation procedure of electroproduction 
amplitudes\cite{pilkuhn}.
Particularly, each of the amplitudes $T_{\eta N \to \pi^0 N}$, 
$T_{\sigma N \to \pi^0 N}$, 
$T_{\rho N \to \pi^0 N}$ and $T_{\omega N \to \pi^0 N}$
are far below the physical region  and near the relevant poles, and therefore
can be approximated as the sum of s, u and t pole terms. 
In fact, a t pole contributes to the 
$T_{\sigma N \to \pi^0 N}$ only and as shown bellow it dominates 
the reaction amplitude.

The situation is drastically different for \p exchange because the 
$T_{\pi^0 N \to \pi^0 N}$ is in the physical region close to threshold. 
It is now well established that the on mass-shell amplitude
for the $\pi^0 p \to \pi^0 p$ scattering is very small and in the soft
pion limit is equal to zero\cite{alfaro} near threshold. Based on 
general quantum mechanical constraints\cite{kowalski,amado}, 
the half off mass-shell
$\pi$N scattering amplitude  can be written as the product of the 
on mass-shell amplitude and a real Kowalski-Noyes function\cite{kowalski},
approaching unity in the limit of $q^2 = m^2_{\pi}$. The derivation of 
this function is difficult and model dependent.
In the traditional treatment of offshellness, this function 
corresponds to the form factor of the B meson leg (see diagram 1), 
which slightly enhances the off mass-shell amplitude with respect to the 
on mass-shell amplitude. Hernandez and Oset\cite{hernandez} suggested 
two models that yield rather large enhancement. 
More detailed 
studies\cite{hanhart,park,miller} , 
agree that the enhancement due to offshellness, though differing in sign,
is significant and must be taken into account.

In order to study the influence of \p exchange we evaluate  the \ppp 
cross section by (1) neglecting \p exchange (this correspond to an exact
cancellation between the direct and rescattering term of 
Refs.\cite{park,miller}), and
(2) using the enhanced $\pi$N amplitude of Park et al.\cite{park}. 
In agreement with Refs.\cite{hanhart,park,miller}, it is found that,
using the off mass-shell rescattering
amplitude  rather than the on mass-shell amplitude, influences the calculated
cross section for the $pp \to pp \pi^0$ rather little.

As noted already, a significant contribution to the 
$T_{\sigma N \to \pi^0 N}$ amplitude is due to the t pole term. 
To calculate this require knowledge of
the $g_{\sigma \pi \pi}$ coupling constant. Taking the Lagrangian density
as $L = g_{\sigma \pi \pi} M {\bf \pi}\cdot {\bf \pi} \sigma$ and
by a simple one 
loop calculations
this constant can be related to the $g_{\sigma NN}$ through, 
\begin{equation}
\frac { g^2_{\sigma \pi \pi}}{4 \pi} 
\approx \left(\frac {4\pi}{3}\right)^2
\frac { g^2_{\sigma NN}}{4\pi} 
\left(\ln {\frac {M}{m_{\pi}}}-1\right)^{-2}
\left(\frac { g^2_{\pi NN}}{4\pi}\right)^{-2}\ +\ 
O\left(\frac { m^2_{\pi}}{M^2}\right)\ \ ,
\label{eq:9}
\end{equation}
where $m_{\pi}$ and $M$ are the pion and proton mass.
With the constants $g^2_{\sigma NN}/4\pi = 8.28$ and 
$g^2_{\pi NN}/ 4\pi = 14.6$ one obtains
$g_{\sigma \pi \pi} = (\ 3.3 \pm 0.1\ ) $. The errors is due to neglecting
terms of the order $O \left(\frac { m^2_{\pi}}{M^2}\right)$.
The transition amplitude of the reaction (solid curve) 
and the different meson 
exchange contributions are drawn in Fig. 2.
The $\sigma$ meson exchange amplitude exceeds by far 
any other contribution. The relative phases for $\sigma$ , \r , \o 
exchanges are predicted to
be +1 and add constructively.
The \e exchange amplitude has a negative phase and interfere
destructively, scaling down the transition amplitude to the proper magnitude
required to explain the data. 

Finally, the cross section is calculated from the 
expression,
\begin{equation}
\sigma_{T} = \frac {M^4}{16 (2\pi)^5 \sqrt(s) {\bf p}_1}
\int  \frac {d^3{\bf p}_3} {E_3} \ \frac {d^3{\bf p}_4} {E_4}\ 
\frac {d^3{\bf p}_{\eta}} {E_{\eta}}   
\ | \ Z_{}\ |^2 {\it {S}}p \left(M^{(in)} \ M^{(in)}{}^{\dagger}
\right) \delta {}^4 (p_i -  p_f)  \nonumber\\  ,
\label{eq:10}
\end{equation}
where $\it {S}p \left(M^{(in)} \ M^{(in)}{}^{\dagger}\right)$ 
denotes the trace
over spinor states, and $Z$ is the three-body FSI correction factor of
Refs.\cite{moalem1, moalem2}.
In the analysis presented below this factor is estimated 
from $\pi$N and NN elastic S-wave scattering phase 
shifts\cite{moalem1,moalem2}. The S-wave
NN phase shift is obtained from the effective range expansion which includes
Coulomb interaction between the two protons. We have used the scattering
length $a_{pp} = -7.82$ fm and an effective range $r_{pp} = 2.7$ fm 
of  Ref.\cite{noyes}. The S11 and S13 $\pi$N  scattering 
lengths are taken to be $a_1 = 0.173 \ m^{-1}_{\pi}$
and $a_3 = -0.101 \ m^{-1}_{\pi}$, respectively\cite{pion}.

Our predictions for the total cross section are drawn 
in Figs. 3 along with the available \ppp data. 
The measured cross-section
is reproduced remarkably well with the \p exchange contribution taken to be
zero (solid curve). Note that all of the model parameters are determined 
independently from NN and $\pi$N scattering data and that none of these
parameters have been adjusted to the \ppp reaction.
Our predictions with the off mass-shell amplitude 
$T_{\pi^0 p \to \pi^0 p}$ of Ref.\cite{park}
are drawn as a dashed curve. As the relative phase of the \p exchange 
contribution and the other meson exchanges is not known we have repeated 
these calculations with the $T_{\pi^0 p \to \pi^0 p}$ taken with
a  reversed sign (small dashed line). 
Thus the effects of the \p virtuality vary
the calculated cross section
by $\approx 15\%$ only, and as in Refs.\cite{hanhart,park,miller} we may
conclude that offshellness of the exchanged \p would not resolve the
discrepancy with experiment.

In agreement with the model of Horowitz 
et al.\cite{horowitz}, the main contribution is due to $\sigma$
exchange with ratios 
$M_{\sigma} : M_{\omega} : M_{\eta} : M_{\rho}
\   \approx \ 100:40:8:7$. Although the contribution
of the $\eta$ meson is rather weak
it becomes effective 
through interference terms with the other exchange contributions. 
It is interesting to note that for the $\sigma$ exchange, 
the sum of the s and u pole terms alone
amounts to $\approx 6\%$ of the \e amplitude only, and that the main 
contribution is due to the t pole term which was not included in previous
studies.
The contribution from the $\delta$ meson 
exchange is extremely weak for two reasons. The
first is that the s and u pole terms have about the same magnitude but
opposite signs. The second is that the mass of the $\delta$ meson is high
and the coupling constant $g_{\delta NN}$ is small. All these scale the 
$\delta$ exchange contribution to less than 0.7$\%$ of the $\eta$ exchange
contribution.

The transition amplitude is practically constant near threshold and
the energy dependence is determined almost solely by the FSI 
correction factor and phase space. 
An important property of the coherent formalism is that the different
two body interactions among the out going particles contribute coherently.
Although the meson-nucleon interactions are weak with respect to the NN
interaction, they become influential through interference. To see this,
we draw in Fig. 4 the cross section corrected for FSI interactions assuming
pure I=$\frac {1}{2}$ and I=$\frac {3}{2}$ interactions for the outgoing
$\pi^0$N pairs. The energy dependence of the cross section differ
significantly for the two channels. But, when
the $\pi$N interactions are taken in the proper 
isospin combination, their overall contribution to the FSI factor almost 
cancels out, leading to an energy dependence practically identical with the
one corrected for the pp \fsi (large dash curve) only.

In summary, we have used OBE model to calculate S-wave production  
cross section for the \ppp reaction by assuming a two-step mechanism
where, a boson formed on one of the nucleons converts into a \p on the
other through the $B N \to \pi^0 N$ process. The kinematic conditions of 
the \ppp reaction at threshold, force the $T_{BN \to \pi^0 N}$ 
amplitudes for the heavy mesons being far 
below the physical region, and hence could be estimated as the sum of
s, u and t pole terms. It is found that the t pole term of $\sigma$
meson exchange is essential in order to reproduce the scale of the 
cross section. Using off mass-shell $\pi$N scattering amplitude as
obtained from chiral perturbation theory would not affect these
conclusions.

\vspace{1.5 cm}
{\bf Acknowledgments} This work was supported in part 
by the Israel Ministry Of Absorption.
We are indebted to  Z. Melamed
for assistance in computation.

\newpage
\vspace{0.4cm}
\begin{figure}
\vspace{4.5in}
\includegraphics{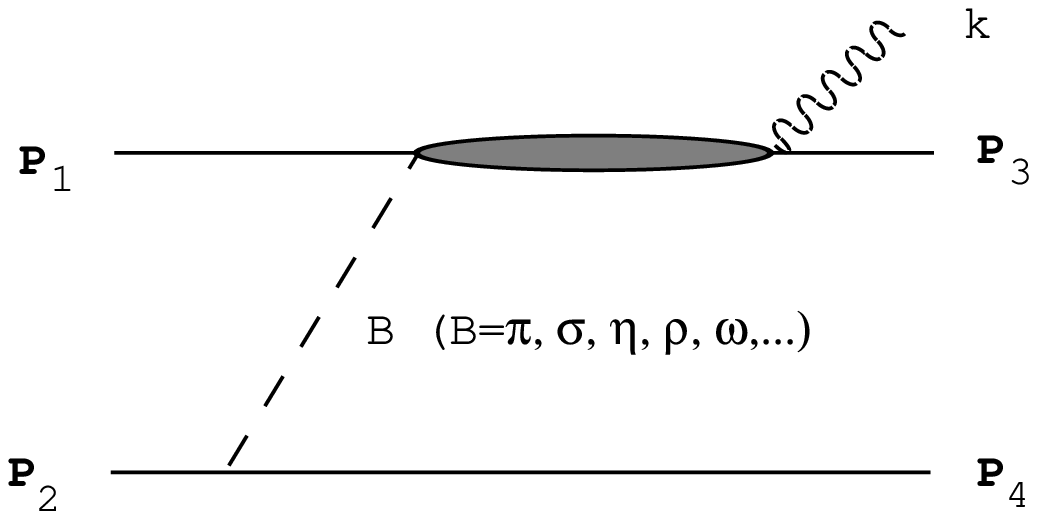}
\vskip 0.5 in
\caption{ The primary production mechanism for the 
$NN \to NN \pi^0$ reaction. 
}  
\end{figure}

\begin{figure}[t]
\vspace{6.0in}
\includegraphics{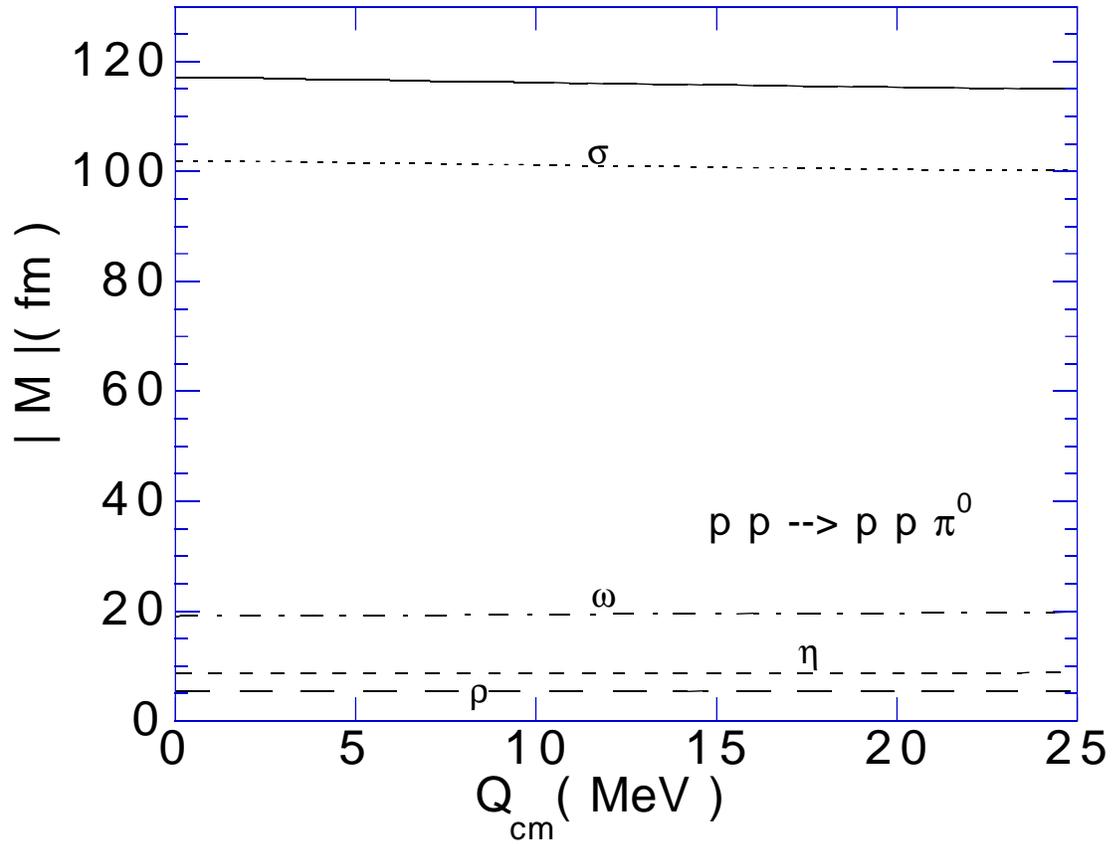}
\vskip 0.5 in
 \caption{  Predictions for the different meson exchange amplitudes 
to the $pp \to pp \pi^0$ reaction vs. $Q_{cm}$, 
the energy available in the center of mass system.
The primary production amplitude is drawn as a solid curve.
Note that the relative phases are taken into account.
}
\label{amplitudes}
\end{figure}

\begin{figure}[t]
\vspace{6.0in}
\includegraphics{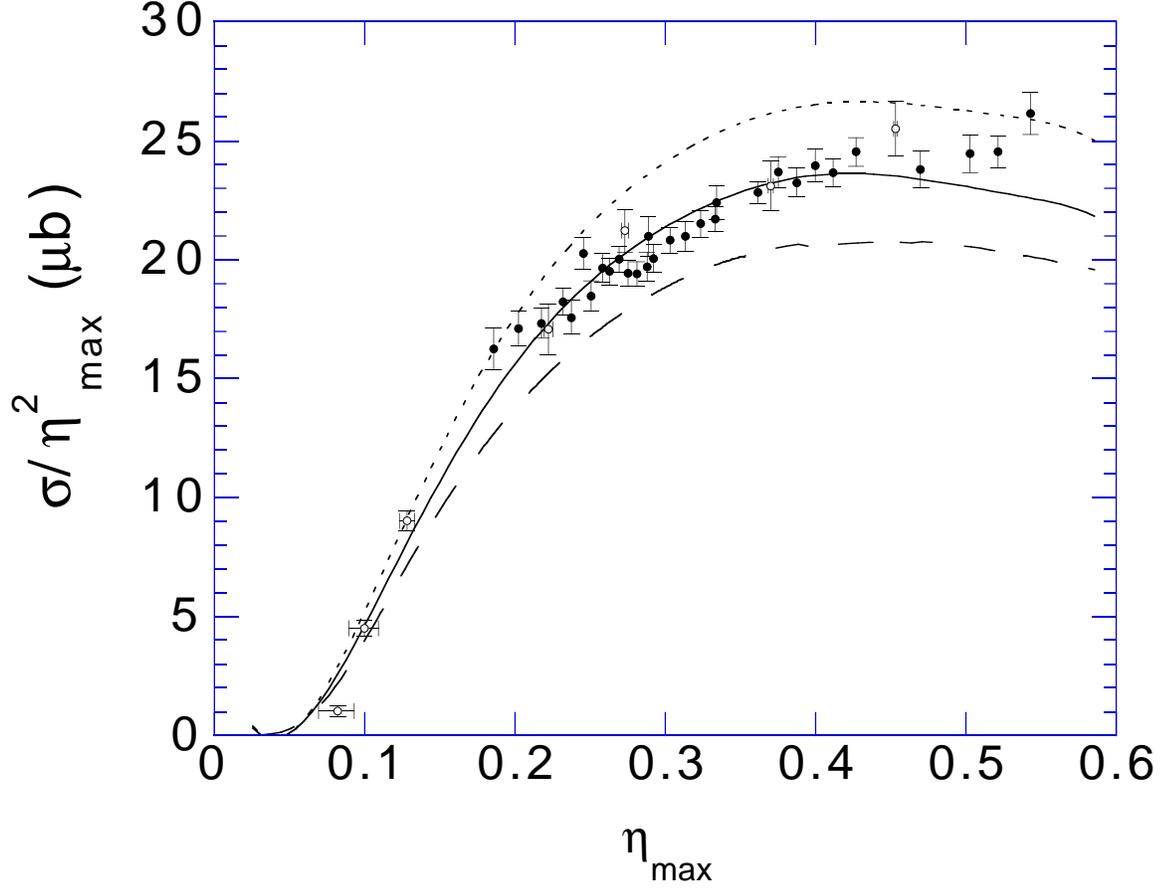}
\vskip 0.5 in
 \caption{  Predictions for the total cross section of the \ppp reaction. 
Predictions with the \p exchange scattering amplitude taken to be zero are
drawn as a solid curve. Those with the $T_{\pi^0 p \to \pi^0 p}$
amplitude taken from Park et al.\cite{park} are drawn as a dashed line. 
The small dashed curve
gives predictions with the amplitude of Park et al.\cite{park}
taken with the sign reversed.
The data points are taken from Refs.\cite{bondar,meyer}
}
    \label{amplitudes}
\end{figure}

\begin{figure}[t]
\vspace{6.0in}
\includegraphics{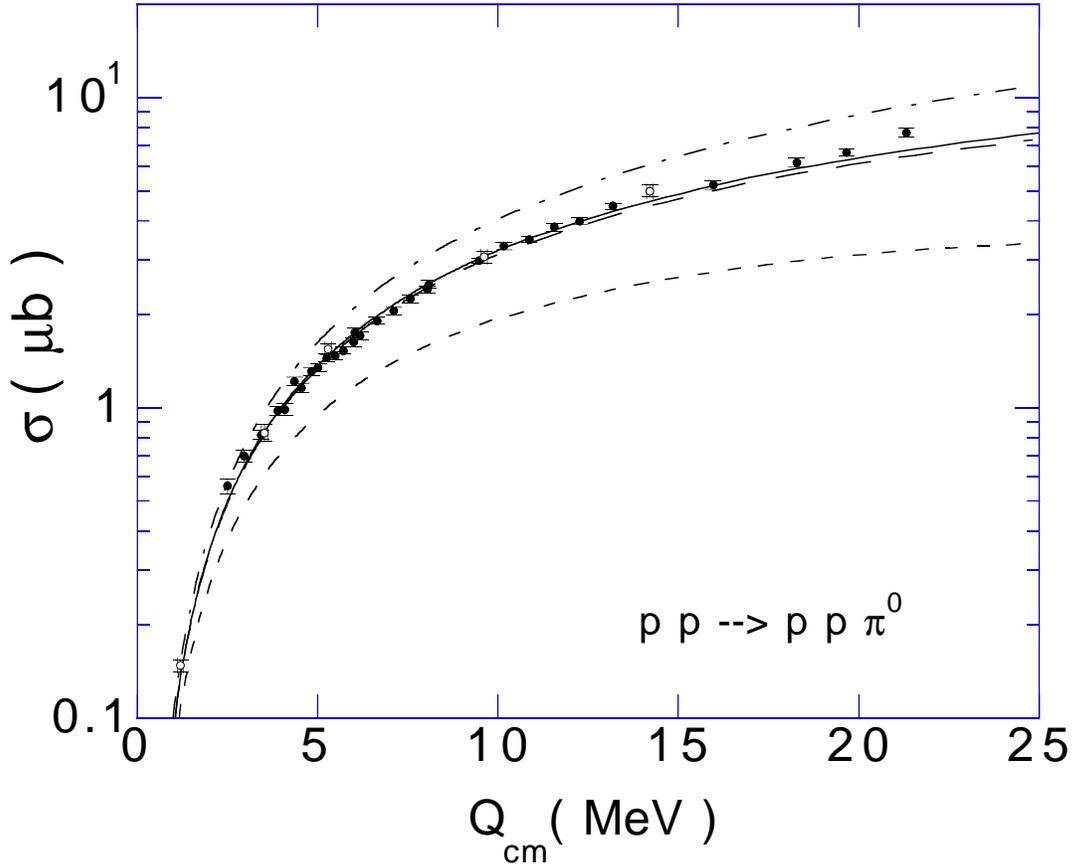}
\vskip 0.5 in
 \caption{ FSI corrections in pure isospin $\pi$N channels.
Integrated energy cross sections calculated with the assumption that 
the interacting $\pi$N pair is scattered in isospin I=$\frac {1}{2}$
(small dashed curve) and I =$\frac {3}{2}$ (dot-dashed curve). 
The solid line is that obtained with the \p N interactions
taken in the appropriate isospin combinations. Predictions which account
for the pp \fsi only are (large dashed curve) are nearly
identical with the solid curve.
}
    \label{iveciscal}
\end{figure}

\end{document}